\documentclass[twocolumn,aps,prb,reprint,groupedaddress]{revtex4}
\usepackage{graphicx}
\usepackage{color}

\newcommand{\be}{\begin{equation}}
\newcommand{\ee}{\end{equation}}

\newcommand{\bea}{\begin{eqnarray}}
\newcommand{\eea}{\end{eqnarray}}

\newcommand{\p}{\partial}

\newcommand{\lb}{\left[}
\newcommand{\rb}{\right]}

\renewcommand{\Im}{{\rm \, Im\,}}
\renewcommand{\vec}[1]{{\bf #1}}
\renewcommand{\hat}[1]{{\widehat #1}}

\begin{document}
\title{Shockley-Ramo theorem and long-range photocurrent response in gapless materials}

\author{Justin C. W. Song$^{1,2}$}
\author{Leonid S. Levitov$^1$}
\affiliation{$^1$ Department of Physics, Massachusetts Institute of Technology, Cambridge, Massachusetts 02139, USA}
\affiliation{$^2$ School of Engineering and Applied Sciences, Harvard University, Cambridge, Massachusetts 02138, USA}





\begin{abstract}
Scanning photocurrent maps of gapless materials, such as graphene, often exhibit complex patterns of hot spots positioned far from current-collecting contacts. We develop a general framework that helps to explain the unusual features of the observed patterns, such as the directional effect and the global character of photoresponse.  
We show that such a  response is captured by a simple Shockley-Ramo-type approach. 
We examine specific examples and show that the photoresponse patterns can serve as a powerful tool to extract information about symmetry breaking, inhomogeneity, chirality, and other local characteristics of the system. 
\end{abstract} 

\maketitle

\section{Introduction}

Many existing schemes of 
photodetection rely on transforming photon energy into electrical signals\cite{sze}. Photoresponse proceeds in three stages: 1) incoming radiation creates electron-hole pairs; 2) photoexcited pairs generate electric fields and charge movement in the system, inducing current in current-collecting contacts; 3) the induced current is  amplified and converted to the output signal. Studies of photogalvanic effects typically concerned with stage 1, focusing on the phenomena occurring locally in the photoexcitation region (see e.g. Refs.\onlinecite{mele,ivchenko,moore,hosur,rana}). 
In contrast, stage 2 received relatively little attention. Here we discuss signal transduction in the system at stage 2, in particular
the mechanisms of spatially non-local response. 

As we will see, these mechanisms have much in common with the processes in charge detectors studied a long time ago by Shockley and Ramo in the context of vacuum-tube electronics.\cite{shockley,ramo,he} They pointed out that the response of charge detectors is governed by long-range effects: the instantaneous electric currents induced by a  moving charge are due to the electric field flux seen by each electrode rather than the amount of charge entering the electrode per second. As a result, the induced currents are only weakly sensitive to the charge position but depend strongly on the charge velocity magnitude and direction. The Shockley-Ramo (SR) approach---the seminal SR theorem---allows one to easily calculate the response. As we demonstrate, even though photoresponse in gapless materials originates from very different physics, it is described by a formalism similar to that of the SR theorem. 

Spatial nonlocality of optoelectronic response is common for many gapped materials where it 
arises due to 
slow recombination of photoexcited carriers\cite{sze}. Recently, however, a long-range photocurrent response was reported in systems where carrier recombination is fast on carrier diffusion timescales. 
Notably, this is the case in scanning photocurrent experiments that probe new gapless materials, such as graphene and topological insulators\cite{bonaccorso,xia,park,lemme,nazin,xu,mciver}. Photoresponse in these systems is of a global character: rather than being localized near current-collecting contacts, the photocurrent hot spots feature complex spatial patterns spanning the entire system area, typically separated by many microns from the contacts\cite{xia,park,lemme,nazin,xu}. These large length scales may seem hard to reconcile with the short picosecond-scale recombination times over which the photoexcited carriers lose their energy and become part of the thermal distribution, traversing distances much less than system size.

The observed  photoresponse also displays other striking features, in particular the {\it directional effect} (Fig.\ref{fig1}). Namely, the photocurrent hot spots are highly sensitive to 
the orientation of inhomogeneities and interfaces at which the hot spots are pinned, 
while being essentially independent of the distance from the contacts. The global character of photoresponse and its strong dependence on the orientation relative to contacts is particularly striking in the data  from Ref.\onlinecite{park} where this effect was first reported
[reproduced in Fig.\ref{fig1}(d)]. Here we introduce a framework that naturally explains how the 
nonlocality can arise in the absence of slow recombination. This framework also provides a simple explanation for the directional effect. 

\section{The origin of the nonlocal and directional behavior}
\label{sec2}

Ambient carriers in gapless materials play an important role in mediating electric currents and transporting energy across the system. Here we analyze long-range photoresponse mediated by such carriers. The reasons the contribution of ambient carriers to photoresponse overwhelms that of primary photoexcited carriers can be summed up as follows. On one hand, short recombination times lead to a rapid decay of the primary photoexcited carriers, preventing them from reaching contacts and directly contributing to photocurrent. On the other hand, ambient carriers can generate currents and fields reaching far from the photoexcitation spot. The main contribution to photoresponse is therefore an indirect one: a local photocurrent sets up an electric field that drives ambient carriers outside the excitation region and into the contacts.

There are several mechanisms by which primary photoexcited carriers can produce local photocurrents in the excitation region. These currents can be due to photovoltaic effects (electron-hole separation by built-in fields) or due to thermoelectric effects. Photovoltaic mechanisms tend to dominate in systems with strong built-in fields (such as semiconductor p-n junctions), whereas thermoelectric mechanisms are important in systems where electron-lattice cooling is slow (such as graphene). The two mechanisms depend on very different length and time scales set by system inhomogeneity, the scattering and recombination mean free paths for photoexcited carriers, the cooling times and lengths for secondary hot carriers, etc. Here, we will not discuss local photocurrent mechanisms in detail, so as to no obscure the main point of this article: on large scales far from the excitation region the response is of a nonlocal Shockley-Ramo type and is mediated by electric currents due to ambient carriers.

\begin{figure}
\includegraphics[scale=0.4] {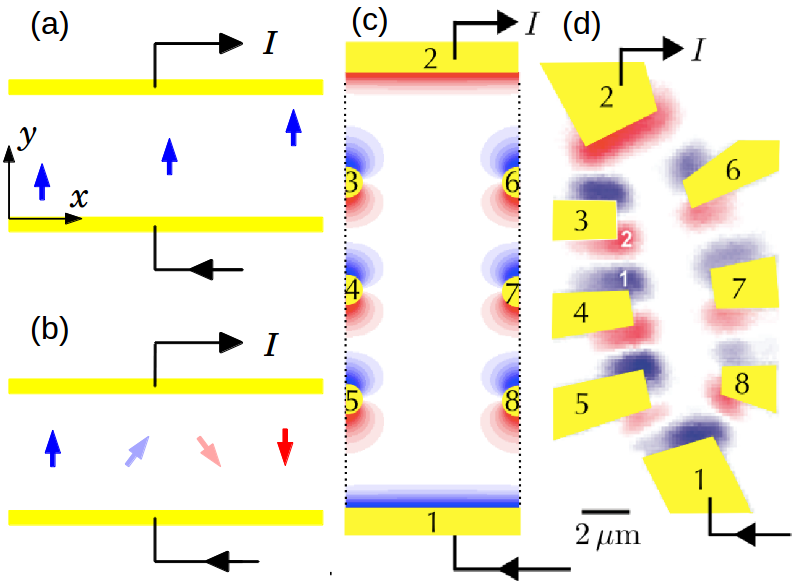} 
\caption{[(a) and (b)] Toy model for long-range photoresponse and directional effect in a strip $0<y<w$ with current-collecting contacts at the sides $y=0,w$ (see Sec. \ref{sec2}). Different photocurrent sources $\vec j_{\rm ph}$ are  schematically shown by arrows. The arrow color and intensity indicate the sign and magnitude of the induced net current $I$. The value $I$ {\it does not depend} on the source position within the strip (a) but has strong dependence on its orientation (b). 
(c) Photocurrent pattern due to floating contacts that do not draw current (yellow semi-circles labeled $3$-$8$). 
The photocurrent, drawn from contacts $1$ and $2$, is modeled as described in Sec. \ref{sec4}, see Eqs.(\ref{eq:conformal_flat}) and (\ref{eq:n(r)}). 
(d) Scanning photocurrent image of a 12 ${\rm \mu m}$-long graphene device with six floating contacts $3$-$8$. Note that the sign of photoresponse near floating contacts is correlated with the direction to the current-collecting contacts $1$ and $2$, but essentially independent of contact location within the system (data taken from Fig.2(a) of Ref. \onlinecite{park}).
}
\vspace{-5mm}
\label{fig1}
\end{figure}

We note that the mechanism discussed here is not the only one that may lead to a long-range photoresponse. For example, in systems with large cooling lengths (such as pristine graphene) hot carriers generated in the excitation region can diffuse across the entire system and reach contacts. Thermopower induced through contact heating by such carriers may create an additional long-range photocurrent response. However, the qualitative features of such photoresponse are quite different from those expected for the Shockley-Ramo-type response. In particular, the number of hot carriers reaching contacts sharply increases when they are excited in proximity to the contacts. Hence, we do not expect the direct heating of contacts to yield a ``global'', position-independent photoresponse. Likewise, since hot carrier generation in the excitation region has no directional dependence, this mechanism alone cannot account for the directional effect which is naturally explained by a Shockley-Ramo-type response.

The processes  can be modelled by a spatially localized ``extraneous'' photogalvanic current $\vec j_{\rm ph}(\vec r)$ induced by photoexcitation, and a diffusion current $\vec j_{\rm d}(\vec r)$ due to ambient carriers in the material, obeying
\be\label{eq:transport}
\nabla \cdot (\vec j_{\rm d}+\vec j_{\rm ph})=0
,\quad
\vec j_{\rm d}=-\sigma (\vec r)\nabla\phi
,
\ee
where $\sigma (\vec r)$ is position-dependent conductivity tensor, $\phi$ is the electrochemical potential. 
As we will see, the resulting response does not diminish with distance and displays the directional effect.

The origin of such a behavior can be understood by 
analyzing a special case: a spatially uniform system with constant conductivity. 
With regard to this toy model, some points of clarification are in order.
First, on general symmetry grounds, local inhomogeneities, interfaces and boundaries are essential for generating photocurrent. Thus, a ``spatially uniform system'' assumption 
only pertains to transport properties far outside the area where $\vec j_{\rm ph}$ is concentrated. 
Second, the assumption of spatial uniformity is used here merely to simplify the discussion. A more general situation will be analyzed in Sec. \ref{sec3}.
Third, as we discuss in Sec. \ref{sec4}, photocurrent patterns are sensitive to the symmetries which govern photoresponse via a relation between $\vec j_{\rm ph}$ and local density gradients, see Eq.(\ref{eq:n(r)}) and accompanying discussion.

As a warm-up, we consider transport in an infinite 2D system in the presence a spatially localized photogalvanic current $\vec j_{\rm ph}(\vec r)$. 
Fourier-transforming transport equations, Eq.(\ref{eq:transport}), yields algebraic equations, giving a non-local relation
\bea\label{eq:D_ik}
&& j_{{\rm d}, i} (\vec r)=\int d^2 r' D_{ik}(\vec r,\vec r')j_{{\rm ph},k} (\vec r')
,\\ \label{eq:D_ik_function}
&& D_{ik}(\vec r,\vec r')=-\sum_q e^{i\vec q(\vec r-\vec r')}\frac{q_i q_k}{\vec q^2}=\frac{2n_in_k-\delta_{ik}}{2\pi(\vec r-\vec r')^2}
,
\eea
where $\vec n$ is a unit vector pointing from $\vec r'$ to $\vec r$. The response function $D_{ik}(\vec r, \vec{r'})$ features strong nonlocality and a directional effect, which are manifest in its power-law decay and angular dependence. 

In writing Eq.(\ref{eq:transport}) we make the usual assumptions that magnetic effects are negligible and the electric field propagates instantaneously. Under these assumptions, the problem can be treated as electrostatics at each moment of charge movement
(with the cutoff frequency value set by the retardation effects due to charge dynamics, see Eq.(\ref{eq:tau}) below).

Next we proceed to demonstrate a relation between the power-law decay found for $D_{ik}$ and the global, position-independent response. We will analyze a simple geometry: a strip $0\le y\le w$ infinite in the $x$ direction, with current-collecting contacts at the sides $y=0,w$, as illustrated in Fig.\ref{fig1}(a,b). We can extend the above analysis to explicitly evaluate  the response induced by a localized source. As we will see, the net current flowing through the contacts equals
\be\label{eq:I_strip}
I=\frac1{w}\int d^2 r'  j_{{\rm ph},y}(\vec r')
.
\ee
This result displays essential nonlocality since $I$ is independent of $\vec j_{\rm ph}$  position [see Fig.\ref{fig1}(a)]. 
While the independence of the $x$ coordinate follows directly from translational invariance, the independence of the $y$ coordinate does not follow from any symmetry. It is counterintuitive and to a large degree comes as a surprise. 


To derive Eq.(\ref{eq:I_strip}), we note that the approach outlined in Eqs.(\ref{eq:D_ik}),(\ref{eq:D_ik_function}) 
can be reformulated in terms of 
the Greens function of Laplace's equation with zero boundary condition at $y=0$, $y=w$,
\be\label{eq:D_strip}
 D_{ik}(\vec r,\vec r')=-\nabla_i G(\vec r,\vec r')\nabla'_k
,\quad 
 \nabla^2 G(\vec r,\vec r')=\delta(\vec r-\vec r')
,
\ee
where $\nabla$ and $\nabla'$ are gradients with respect to $\vec r$ and $\vec r'$. 
 Fourier-transforming with respect to $x$, we express the result through a 1D Greens function, $G(\vec r,\vec r')=\sum_q e^{iq(x-x')}g_q(y,y')$, 
\be
(\p_y^2-q^2) g_q(y,y')=\delta(y-y')
.
\ee
Solving this equation in the interval $[0,w]$ with zero boundary conditions, we obtain 
\be
g_q(y,y')=A\sinh(qy_<)\sinh q(y_>-w)
,
\ee 
where $y_<={\rm min}\,(y,y')$, $y_>={\rm max}\,(y,y')$, $A=\frac1{q\sinh(qw)}$. Plugging this into Eqs.(\ref{eq:D_strip}) and (\ref{eq:D_ik}) and setting $y=0$, we find the normal current at the boundary, 
$j_n^{\rm (d)}(x)=j_{{\rm d}, y}(x)_{y=0}$. We obtain
\be
j_n^{\rm (d)}(x)=-\int d^2r' \sum_q e^{iq(x-x')} \frac{\sinh q(y'-w)}{\sinh(qw)} 
\nabla'\cdot\vec j_{\rm ph}(\vec r')
.
\ee
By mirror symmetry, only the component of $\vec j_{\rm ph}$ normal to the strip contributes to the above expression. Integration by parts gives
\bea\nonumber
&&\int_0^w dy' \frac{\sinh q(y'-w)}{\sinh(qw)} 
\p_{y'} j_{{\rm ph},y}(y')= j_{{\rm ph},y}(y'=0)
\\
&&-\int_0^w dy' \frac{q\cosh q(y'-w)}{\sinh(qw)}  j_{{\rm ph},y}(y')
\label{eq:eq9}
\eea
The net current 
is evaluated as $I=\int dx (j_{{\rm d}, y}(x)+j_{{\rm ph}, y}(x))_{y=0}$, where the last term cancels with an identical term in Eq. (\ref{eq:eq9}) right-hand side. Using the relation $\int dx e^{iq(x-x')} =2\pi\delta(q) $ we arrive at  the result in Eq.(\ref{eq:I_strip}). In addition to the ``global property'' (independence of $\vec j_{\rm ph}$ position), our result also displays the ``directional property'' since the response depends on the $y$ component of $\vec j_{\rm ph}$ only, reversing sign upon $\vec j_{\rm ph}$ reversal [see Fig. \ref{fig1}(b)]. 


It is instructive to note a relation between our calculation 
above and an electrostatic problem of a point dipole inserted in a parallel plate capacitor. The dipole induces image charges on the capacitor plates, which also  display the directional property and the global property. Namely, the net induced charge values are given by $\Delta q_{1,2}=\pm \frac1{w}p\cos\theta$, where $p$ and $\theta$ are the dipole magnitude and tilt angle, and $w$ is the plate separation. The dependence of $\Delta q_{1,2}$ on $\theta$ and their independence of dipole position are identical to that for photoresponse, as illustrated in Fig.\ref{fig1}(a,b). The origin of this relation can be traced to an isomorphism between the two problems, with $\vec j_{\rm d}$ and $\vec j_{\rm ph}$ playing the role of the electric field and dipole density in the electrostatic problem. 
As we will see in the next section, this result can be viewed as a special case of the SR theorem.

\section{Mapping to the Shockley-Ramo problem}
\label{sec3}

The global property and the directional property 
bear strong resemblance to the behavior in charge detectors described by the SR approach\cite{shockley,ramo,he}. Before working out the connection between our problem and the SR approach, we briefly summarize the key facts. Shockley and Ramo were concerned with the 
currents induced in the electrodes by charges moving in the free space inside a vacuum tube.  
The SR theorem provides a closed-form relation between the current induced by a moving charge $e$ in the electrode $k$ and the charge velocity and position, denoted by  $I_{k}$, $\vec v(t)$ and $\vec R(t)$, respectively. The SR result, which is intrinsically nonlocal due to the long-range character of electric fields in vacuum,  reads
\be\label{eq:SR_canonical}
I_k= e\vec v(t)\cdot\vec E_{\vec r=\vec R(t)},\quad
\vec E(\vec r)=\nabla w_k(\vec r)
,
\ee
The ``weighting potentials'' $w_k(\vec r)$ satisfy Laplace's equation with suitable boundary conditions on the electrodes ($w_k=1$ at electrode $k$, and $w_k=0$ at electrodes $j\ne k$). 
The SR theorem is a foundation of ultra-fast charge sensing, such as
particle detection in high energy physics\cite{yoder,he} and plasma diagnostics.\cite{maero} It can also be extended to charges moving in insulators\cite{cavalleri}.

In contrast, the relation between our problem and the SR treatment of charge detectors constitutes a mapping rather than a direct application of the SR approach. 
In particular, the flow of ambient carriers and the photocurrent source play the role of electric field and moving charge in the SR problem,
respectively. The long-range character of the response can be linked to charge continuity. 
The condition $\nabla \cdot \vec j = 0 $ can be interpreted as incompressibility of current flow, with stream lines that do not terminate anywhere within the system.
In addition, because the current is caused by a chemical potential gradient, the stream lines 
cannot form loops.
This results in a response not diminishing with the distance between contacts
and local photoexcitation, $\vec j_{\rm ph}$.
As we show below, basically following the SR strategy, the system response  can be described as
\be
\label{eq:SR}
I=A\int \vec j_{\rm ph}(\vec r)\cdot \nabla\psi (\vec r)\, d^2r
,
\ee
where $\vec{j_{\rm ph}}(\vec r)$ is the local photogalvanic current in the photoexcitation region, $\psi$ is a weighting field obtained by solving a suitable Laplace problem, and $A$ is a prefactor which depends on device configuration (see Eq.(\ref{eq:A})).

\begin{figure}[t]
\includegraphics[scale=0.25]{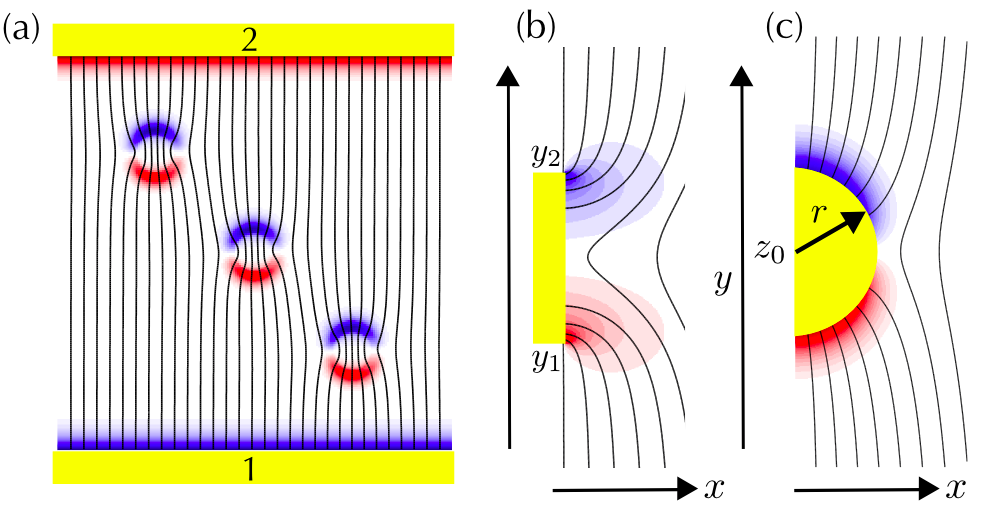}
\caption{Directional effect in photoresponse accounting fully for the distortions of the weighting field.
(a) Photocurrent pattern due to three circular regions, modeled in the same way as in Fig.\ref{fig1B} (b).  The conductivity inside each region is taken to be 10 times larger than the background conductivity. 
(b,c) Photoresponse and the field lines for $\nabla\psi$ near floating contacts of two different shapes, a rectangle and a semicircle, obtained using the conformal mapping approach, Eq.(\ref{eq:conformal_flat}).
}
\label{fig2}
\end{figure}

As illustrated in Figs.\ref{fig1},\ref{fig2},\ref{fig1B}),  Eq.(\ref{eq:SR}) predicts photocurrent-active structures with contrast which is essentially independent on their position within the system. Such ``global'' photoresponse is known for one-dimensional systems, where Eq.(\ref{eq:SR}) reduces to adding up the total potential drop across the device \cite{nelson}. However, the generalized framework presented here yields photocurrent that can exhibit complex structures which are not anticipated in a one-dimensional approach. 

We emphasize that the origin of nonlocality in our photoresponse problem is markedly different from that in the SR problem, since the ambient carriers screen the long range electric field created by photoexcited carriers. 
As noted above, the nonlocality originates from long-range currents constrained by charge continuity relation. 
Further, the SR theorem is typically applied to high-speed charge detection, whereas we are concerned with the steady-state photocurrent. 
Yet, despite these differences, our approach yields a relation [Eq.(\ref{eq:SR})] which exhibits formal similarity with the SR theorem.

The cornerstones of our analysis is the continuity equation, Eq.(\ref{eq:transport}). 
As discussed above, the two contributions to current in Eq.(\ref{eq:transport}) have very different spatial dependence: the photogalvanic current $\vec j_{\rm ph}$ is present in the excitation region, whereas the diffusion current $\vec j_{\rm d}$ is nonzero throughout the entire material.
Below we focus on 
the simplest situation 
when transport can be described by a position-dependent $2\times 2$ conductivity tensor $\sigma (\vec r)$.
The diffusion current satisfies the usual local relation $\vec j_{\rm d}=-\sigma (\vec r)\nabla\phi$. 
The boundary conditions in this transport problem are zero current through the sample boundary, $\vec n\cdot(\vec j_{\rm d}+\vec j_{\rm ph})=0$, and constant potential at the contacts, $\vec n\times\nabla\phi=0$ (here $\vec n$ is the normal to the boundary).

To handle the non-local response, we introduce an auxiliary weighting field 
$\psi(\vec r)$  in the bulk of the material, satisfying 
\be
\nabla \cdot \vec j^{(\psi)}(\vec r)=0
,\quad
\vec j^{(\psi)}=-\sigma^{\rm T}\nabla\psi
,
\ee 
where $\sigma^{\rm T}$ is a $2\times 2$ matrix transposed to $\sigma$, and $\vec j^{(\psi)}(\vec r)$ is an auxiliary current density. The fields $\psi(\vec r)$, $\vec j^{(\psi)}(\vec r)$ satisfy
natural boundary conditions at the boundary and contacts, $\vec n\cdot \vec j^{(\psi)}(\vec r)=0$ and $\vec n\times\nabla\psi(\vec r)=0$, respectively (here $\vec n$ is a normal unit vector at the boundary). Multiplying the continuity equation for the physical current $\vec j_{\rm d}+\vec j_{\rm ph}$ by $\psi(\vec r)$, integrating over the sample area, and using Gauss' theorem, we obtain
\be\label{eq:SR_general}
\int \nabla\psi(\vec r)\cdot\vec j_{\rm ph}(\vec r) d^2r=\sum_k \psi_k I_k-\phi_k I_k^{(\psi)}
\ee
where $k$ labels contacts. The quantities on the right hand side are the net currents flowing in each of the contacts, $I_k=\int_{C_k}\vec n\cdot\vec j_k d\ell$, and potentials on these contacts. We emphasize that Eq.(\ref{eq:SR_general}) holds on very general grounds regardless of whether a particular contact is drawing current ($I_k\ne 0$) or is floating ($I_k=0$). The expression on the left hand side depends on the microscopic distribution $\vec j_{\rm ph}(\vec r)$ inside the material, whereas the expression on the right hand side is a function of currents and potentials at the contacts, thereby providing a general relation between position-dependent photoexcitation and the measured photocurrent.

It is convenient to choose $\psi(\vec r)$ such that $I_k^{(\psi)} = 0$ for all floating contacts. Then the contribution to Eq.(\ref{eq:SR_general}) due to floating contacts drops out entirely, yielding a relation which only includes the contacts that actually draw current. It is also straightforward to account for the effect of an external circuit. We consider the current drawn through a pair of contacts $1$ and $2$ (see Fig.\ref{fig1}) and write $I_{1(2)}^{(\psi)}=\mp(\psi_1-\psi_2)/R$, $I_{1(2)}=\pm(\phi_1-\phi_2)/R_{\rm ext}$, with $R$ and $R_{\rm ext}$ the resistance of the sample and of the external circuit, respectively. Setting $\psi_1-\psi_2=1$, we obtain Eq.(\ref{eq:SR}) with the prefactor
\be\label{eq:A}
A = R/(R+ R_{\rm ext}).
\ee 
Despite its apparent simplicity, Eq.(\ref{eq:SR}) accounts for all the key effects that impact photoresponse, such as system geometry, structure, inhomogeneity, and so on. Similarly to the canonical SR relation, Eq.(\ref{eq:SR_canonical}), the relation in Eq.(\ref{eq:SR}) is essentially  nonlocal due to the long-range character of currents in the system.

Next we briefly discuss the validity of our approach. 
Our transport equations, Eq.(\ref{eq:transport}), are written in a quasistatic approximation. This is similar to the SR approach which treats the electric field induced by a moving charge as instantaneous. The SR result is therefore valid at frequencies below the cutoff set by the EM retardation timescale, $\omega\ll\omega_0=c/L$, where $L$ is system size. In our case, the cutoff frequency is set by the characteristic time for charge dynamics in the system. 
An estimate below yields very short timescales, i.e. a very fast response.

A crude estimate of timescales can be obtained by reinstating the time dependent term in the continuity equation. For a spatially uniform system, the dynamics of the Fourier harmonics of charge density is given by
\be
\p_t \delta n_{\vec k}(t)=-\frac{2\pi}{\kappa} \sigma |{\vec k}|\delta n_{\vec k}(t)
,
\ee
where $\sigma$ is the sheet conductivity per square area
and $\kappa$ is the dielectric constant. For a simple estimate, taking parameter values $|{\vec k}|\approx \pi/L$, $L=10\,{\rm \mu m}$, $\kappa=5$, $1/\sigma=1\,{\rm k\Omega}$, we obtain a sub-picosecond response time
\be\label{eq:tau}
\tau= \kappa L/(2\pi^2\sigma)
\approx 0.3\,{\rm ps},
\ee
which is considerably shorter than typical cooling and recombination times in graphene. 
Fast response makes the photocurrent a potentially useful probe for the dynamical processes in the excitation region. It also makes gapless materials viable for applications in high-speed optoelectronics.

\section{Geometry of the weighting field}
\label{sec4}

The general features of Eq.(\ref{eq:SR}) can be illustrated for a spatially uniform system of a rectangular shape. In this case, the weighting field $\psi(\vec r)$ is a linear function, $\nabla \psi  = \hat{\vec y} /L$, with $L$ the system length.
Constant $\nabla \psi$ yields Eq.(\ref{eq:I_strip}) derived in Sec. \ref{sec2} by a direct calculation. As discussed above, this describes
a response which is invariant upon spatial translation of $\vec j_{\rm ph}(\vec r)$ (the global property). At the same time, the sign and the magnitude of the response depend on the angle between $\nabla \psi (\vec r)$ and $\vec{j_{\rm ph}}(\vec r)$ (the directional effect).

To test the robustness of the global and directional effects, we now proceed to analyze a more realistic situation where spatial inhomogeneity in conductivity $\sigma(\vec r)$ is essential. 
In this case, we use a numerical procedure to obtain the exact profile $\psi(\vec r)$. Fig.\ref{fig2}(a) shows photocurrent patterns from three circular regions with a mismatch between the inner and outer conductivity, 
which causes significant distortions of the $\nabla\psi$ field lines. Yet these distortions do not impact the overall trends discussed above, the global character of the response and the directional effect. This is manifest in the identical spatial structures of the response for all three patterns shown in Fig. \ref{fig2}(a), in both the photocurrent intensity and its angular distribution.

Interestingly, the weighting field distortions have a very dramatic effect near contacts. Even if a contact does not draw net current, it {\it short-circuits} the current flowing in its vicinity, leading to a non-vanishing normal component of $\nabla\psi$ near the surface of a contact (see Fig. \ref{fig2}).
For $\vec{j_{\rm ph}}$ which is normal to the contact, this gives a nonzero, sign-changing photoresponse, as in Fig.\ref{fig1} (c,d).

For ideal contacts, the field $\psi$ can be found using the conformal mapping approach, giving $\psi(\vec r)=A\Im w(z)$. Here $w$ is a suitable analytic function of a complex variable $z=x+iy$, which satisfies the equipotential condition at the contact surface.
We illustrate this for a flat contact and for a semicircular contact (see Fig.\ref{fig2} (b,c)): 
\be\label{eq:conformal_flat}
w_{\rm b}(z)=\sqrt{(z-y_1)(z-y_2)}, \quad 
w_{\rm c}(z) = \tilde  z - r^2/\tilde z,
\ee
$\tilde z=z-z_0$, where the flat contact is positioned at $y_1<y<y_2$, $x=0$,
and the semicircular contact is of radius $r$ and is positioned at $z=z_0$. We assume that the contacts are floating and are small compared to the system size. 
At large $z$, $\psi$ asymptotically approaches the linear dependence $\psi\propto y$ found above. 
The photocurrent at the contact is proportional to $\vec n\cdot\nabla\psi$. For the flat contact, 
\be\label{eq:p_y_psi}
\p_x\psi(\vec r)_{x=0}=
A\frac{y-\frac12(y_1+y_2)}{\sqrt{(y-y_1)(y_2-y)}}
,\quad y_1<y<y_2
.
\ee
Since this quantity is an odd function of $y-\frac12(y_1+y_2)$, the net current drawn in the contact vanishes, 
as appropriate for a floating contact. 
Similar sign-changing behavior is found for the semicircular contact, see Fig.\ref{fig2} (c). The sign-changing pattern is oriented in such a way that the parts showing high photoresponse are facing the contacts through which the photocurrent is drawn. This behavior is in agreement with the directional effect, see Fig.\ref{fig1}(d). 


\begin{figure}
\includegraphics[scale=0.26]  {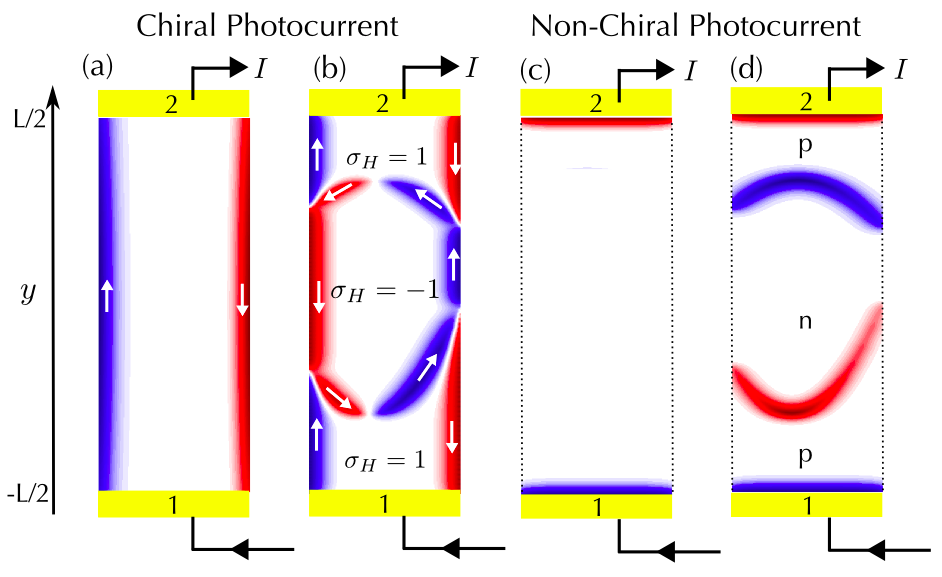} 
\caption{Scanning photocurrent images for different mechanisms of photoresponse. 
The photocurrent, drawn from contacts $1$ and $2$, is modeled by Eqs.(\ref{eq:SR}),(\ref{eq:localpc}). (a,b) Photocurrent pattern in a homogeneous chiral material (a), and inhomogeneous chiral material (b) where $\sigma_H = \pm 1$ marks regions of different chirality. 
Here local photocurrent direction is governed by edge states (white arrows).
(c,d)  Photocurrent pattern in homogeneous nonchiral system (c), and inhomogeneous nonchiral system (d) with a step-like density inhomogeneity (see text). 
}
\vspace{-5mm}
\label{fig1B}
\end{figure}

Next, we discuss application of our approach for diagnostic of different types of photogalvanic response. 
The value $\vec{j_{\rm ph}}(\vec r)$ depends on system properties in the photoexcitation region. By symmetry, no photogalvanic effect can occur in a spatially uniform system (assuming unpolarized light). 
In the presence of a density gradient $\nabla n(\vec r)$, the local photogalvanic current can be described as
\be\label{eq:n(r)}
\vec{j_{\rm ph}}(\vec r) = \lb \alpha \hat{\vec z} \times \nabla n(\vec r)+ \beta \nabla n(\vec{r} ) \rb W(\vec r),
\label{eq:localpc}
\ee
where $\alpha$ and $\beta$ are material constants, and $W(\vec r)$ is the absorbed optical power. In general, 
$\beta$ is finite in all materials, whereas $\alpha$ is only non-zero in {\it chiral systems} where edge-state transport allows $\vec{j_{\rm ph}}$ to be directed along the contours of $n(\vec r)$. 
This is the case in chiral materials such as topological insulators due to coupling between orbital motion and spin \cite{mciver,hosur,moore}, or in non-chiral materials in the presence of a magnetic field\cite{nazin}. 

The effects of spatial inhomogeneity are illustrated in Fig.\ref{fig1B} for the chiral response [$\alpha$ finite, $\beta=0$, see Figs. \ref{fig1B}(a,b)] and a nonchiral response [$\alpha=0$, $\beta$ finite, see Figs. \ref{fig1B}(c,d)]. 
The patterns in Fig.\ref{fig1B} were obtained using a spatially uniform weighting field approximation, $\nabla\psi\approx\hat{\vec y}/L$. For the homogeneous case, Figs. \ref{fig1B}(a,c), we use a constant density $n$ inside the device boundaries and zero density outside. For the inhomogeneous case Figs. \ref{fig1B}(b,d), we use a steplike density profile, with $n$ taking one value in the middle region and another value in the top and bottom regions, identical for Figs. \ref{fig1B}(b) and \ref{fig1B}(d). In both cases, the photocurrent is zero in the regions of constant $n$ and nonzero near the steps. The differences in the sign and magnitude of the response reflect the fundamental difference in physics in the cases shown in Fig. \ref{fig1B}.

Model (a) describes photoresponse in chiral systems arising at 
the interfaces between domains of opposite chirality. Physically, it may represent a quantum Hall system near a plateau transition\cite{chklovskii}, or a system in which nonzero chirality results from spontaneous ordering\cite{spivak}. The different signs of chirality, labelled by $\sigma_H=\pm 1$ in Fig.\ref{fig1B}(a), can be associated with the clockwise and counter-clockwise edge states, labelled by white arrows. Notably, the sign and magnitude of photocurrent depend on the direction of current flow in the edge states.  The photocurrent is also nonzero at system boundaries, indicating the presence of current carrying edge states. This can be used to identify the edge states and domains with different chirality in experiment.

Fig.\ref{fig1B} (a,b) describes photoresponse in chiral systems peaking at the edges of the device for a homogeneous system [Fig. \ref{fig1B}(a)] and arising at the interfaces between domains of opposite chirality [Fig. \ref{fig1B}(b)]. 
Physically, it may represent a quantum Hall system near a plateau transition\cite{chklovskii}, or a system in which nonzero chirality results from spontaneous ordering\cite{spivak}.  The different signs of chirality, labelled by $\sigma_H=\pm 1$ in Fig.\ref{fig1B}(a), can be associated with the clockwise and counter-clockwise edge states, labelled by white arrows. Notably, the sign and magnitude of photocurrent depend on the direction of current flow in the edge states.  In both cases (a,b), the photocurrent is nonzero at system boundaries, indicating the presence of current carrying edge states. This can be used to identify the edge states and domains with different chirality in experiment.

Fig.\ref{fig1B} (c,d) shows nonchiral photocurrent response for a homogeneous [Fig. \ref{fig1B}(c)] and inhomogeneous [Fig. \ref{fig1B}(d)] system. Physically, Fig. \ref{fig1B}(d)  may describe systems such as graphene with spatial inhomogeneity giving rise to p-n boundaries separating regions with electron-like and hole-like polarity \cite{lemme}. In this case, $\vec{j_{\rm ph}}$ is normal to the contours of $n(\vec r)$, making the sign and magnitude of the response dependent on the orientation of the interfaces viz.  $\hat{\vec y}\cdot \vec{j_{\rm ph}}$. Also, since $\vec{j_{\rm ph}}$ is normal to boundaries whereas $\nabla\psi$ is tangential, the photocurrent vanishes at the system edge.

A very different behavior is found near contacts, since $\nabla\psi$ is normal to the contact surface, see Fig.\ref{fig1} (c). In this case, a nonzero response arises both near the contacts through which current is drawn and near floating contacts 
(see also Fig.\ref{fig2}). Notably, the response depends on the floating contact orientation but not on its position within the system. This is in agreement with experimental observations of Ref.\onlinecite{park}, which are reproduced in Fig. \ref{fig1} (d). All photocurrent patterns in Fig.\ref{fig1} and Fig.\ref{fig1B}, despite their different physical origin, share two common trends: strong directional sensitivity and global character (positional independence). 
This behavior makes the photocurrent patterns particularly useful in identifying symmetry breaking and inhomogeneity in gapless materials.

\section{Conclusions}

In summary, our approach explains several puzzling aspects of photocurrent response in gapless materials, in particular
the striking non-locality and the directional effect observed in Ref.\onlinecite{park}. By analyzing 
different mechanisms of photoresponse, we demonstrate that it is uniquely capable of revealing spatial patterns arising due to symmetry breaking, chirality, or inhomogeneities.  
There are several other mechanisms that may conceivably result in a nonlocal photocurrent response. One such mechanism is the nonlocal current-field relation predicted for atomically thin systems in Ref. \onlinecite{rashba}. Another is the nonlocality mediated by charge-neutral modes, such as spin or energy \cite{abaninnonlocal,folk14}. However, we believe that these mechanisms cannot account for the global and directional effects. Our results therefore indicate that system-wide electric currents mediated by ambient carriers constitute the main mechanism responsible for the observed long-range photocurrent response. 

\section{Acknowledgments}

We acknowledge useful discussions with M. Rudner and X. Xu, and financial support from the NSS program, Singapore (J.S.), and the Center for Excitonics, and Energy Frontier Research Center funded by the U.S. Department of Energy, Office of Science, Basic Energy Sciences under Award No. DE-SC0001088 (L.L.).

\end{document}